# Dispelling the myth of robotic efficiency: why human space exploration will tell us more about the Solar System than will robotic exploration alone.

**Ian A. Crawford**
Department of Earth and Planetary Sciences, Birkbeck College London, Malet Street, London, WC1E 7HX (i.crawford@bbk.ac.uk).

**Abstract**
There is a widely held view in the astronomical community that unmanned robotic space vehicles are, and will always be, more efficient explorers of planetary surfaces than astronauts (e.g. Coates, 2001; Clements 2009; Rees 2011). Partly this is due to a common assumption that robotic exploration is *cheaper* than human exploration (although, as we shall see, this isn't necessarily true if like is compared with like), and partly from the expectation that continued developments in technology will relentlessly increase the capability, and reduce the size and cost, of robotic missions to the point that human exploration will not be able to compete. I will argue below that the experience of human exploration during the Apollo missions, more recent field analogue studies, and trends in robotic space exploration actually all point to exactly the opposite conclusion.

**Benefits of human space exploration**

As demonstrated by the Apollo missions forty years ago, and leaving aside the question of cost (which is not straightforward and which I address separately below), human space exploration has a number of advantages over robotic operations on planetary surfaces. These have been discussed in detail elsewhere (e.g. Spudis 2001; Crawford 2001, 2010; Garvin 2004; Cockell 2004; Snook et al. 2007), and were endorsed by the independent Commission on the Scientific Case for Human Space Exploration commissioned by the RAS in 2005 (Close et al., 2005; hereinafter 'the RAS Report'). These advantages may briefly be summarised as follows:

- On-the-spot decision making and flexibility, with increased opportunities for making serendipitous discoveries;

- Greatly enhanced mobility and attendant opportunities for geological exploration and instrument deployment (compare the 35.7 km traversed in three days by the Apollo 17 astronauts in December 1972 with the almost identical distance (34.4 km) traversed by the Mars Exploration Rover *Opportunity* in eight *years* from January 2004 to December 2011);

- Greatly increased efficiency in sample collection and sample return capacity (compare the 382 kg of samples returned by Apollo with the 0.32 kg returned by the Russian robotic sample return missions *Luna*s 16, 20 and 24, and the zero kg returned to-date by any robotic mission to Mars);

- Increased potential for large-scale exploratory activities (e.g. drilling) and the deployment and maintenance of complex equipment; and

- The development of a space-based infrastructure capable of supporting space-based astronomy and other scientific applications (e.g. the construction and maintenance of large space telescopes).

With the exception of the final bullet point, for which the best demonstration is provided by the five successful space shuttle servicing missions to the HST (e.g. NRC 2005), demonstration of the benefits of human spaceflight for planetary exploration must be sought in a comparison of the relative efficiencies of the Apollo missions and robotic missions to the Moon and Mars, supported where appropriate with terrestrial analogue studies.

The relative efficiency of human over robotic exploration of planetary surfaces is in fact well recognized by scientists directly involved with the latter on a day-to-day basis. For example, regarding the exploration of Mars, the RAS Report noted that:

> *"the expert evidence we have heard strongly suggests that the use of autonomous robots alone will very significantly limit what can be learned about our nearest potentially habitable planet"* (Close et al. (2005; paragraph 70).

Putting it more bluntly, Steve Squyres, the Principal Investigator for the Mars Exploration Rovers *Spirit* and *Opportunity*, has written:

> *"[t]he unfortunate truth is that most things our rovers can do in a perfect sol* [i.e. a martian day] *a human explorer could do in less than a minute"* (Squyres, 2005, pp. 234-5).

This is of course only a qualitative assessment, albeit by someone well placed to make an informed judgement. Nevertheless, taken at face value it implies a human/robot efficiency ratio of about 1500, which is far larger than the likely ratio of cost between a human mission to Mars and the cost of the MERs (see below). Even this, however, does not fairly compare human exploration efficiency with robotic exploration. This is because much of the scientific benefit of human missions will consist of samples returned, drill cores drilled and geophysical instruments deployed, all of which were demonstrated by Apollo on the Moon, but none of which have been achieved by the MERs nor will be achieved by the more capable (and vastly more expensive) Mars Science Laboratory (MSL) that is due to land on Mars in 2012.

More objective estimates of the relative efficiency of robots and humans as field geologists have been given by Garvin et al. (2004) and Snook et al. (2007). Garvin (2004) summarised the results of a NASA survey of several dozen planetary scientists and engineers on the relative efficiency of human and robotic capabilities in 18 different skill sets relevant to planetary exploration. The results are summarised in Figure 1, and show a clear balance in favour of human capabilities (with the implicit recognition that the most efficient exploration strategies of all will be those consisting of human-robotic partnerships where each complements the other).

This conclusion is corroborated by direct field comparisons of human and robotic exploration at planetary analogue sites on Earth. Snook et al. (2007) reported the results of one such study, conducted at the Haughton impact crater in the Canadian arctic, where the efficiency of a human explorer (suitably encumbered in a spacesuit) was compared with that of a tele-operated rover (controlled from NASA Ames Research Centre in California) in the performance of a range of exploration tasks. The rover was more sophisticated than those employed in present-day space missions, and included simulation of artificial intelligence capabilities that are only likely to be incorporated in actual space missions from 2015 at the earliest. Nevertheless, the space-suited 'astronaut' was found to be much more efficient in performing exploration tasks than the rover, and Snook et al. (2007; p. 438) concluded that: "humans could be 1-2 orders of magnitude more productive per unit time in exploration than future terrestrially controlled robots."

Although this estimate is an order of magnitude lower than Squyres' off-the-cuff estimate of 1500 given above, this is mainly because the comparison was conducted between human and *tele-robotic* exploration, rather than between humans and supervised quasi-autonomous robotic exploration such as carried out by the MERs and MSL. Tele-robotic exploration is known to be more efficient than autonomous robotic operation, precisely because real-time human interaction is involved, but it cannot be employed effectively on planetary surfaces more distant than the Moon owing to the inevitable communications delay (Lester and Thronson 2011). Garvin (2004, see his Fig. 2) has compared the efficiencies of robotic, tele-robotic, and human exploration, from which it is clear that if humans are '1-2 orders of magnitude more efficient' than tele-robots then they will be even *more* efficient when compared with robotic vehicles like the MERs or MSL, bringing the two estimates into better agreement.

Moreover, as noted above, comparisons based on the relative time taken to perform certain tasks, while they do indeed show humans to be more efficient than robots, nevertheless still grossly underestimate the added scientific value of having humans on planetary surfaces. This is because astronauts have to come back to Earth, and can therefore bring large quantities of intelligently collected samples back with them. Robotic explorers, on the other hand, generally do not return (this is one reason why they are cheaper!) so nothing can come back with them. Even if robotic sample return missions are implemented, neither the quantity nor the diversity of these samples will be as high as would be achievable in the context of a human mission -- again compare the 382 kg of samples (collected from over 2000 discrete locations) returned by Apollo, with the 0.32 kg (collected from three locations) brought back by the *Luna* sample return missions. The Apollo sample haul might also be compared with the ≤0.5 kg generally considered in the context of future robotic Mars sample return missions (e.g. ISAG, 2011). Note that this comparison is not intended in any way to downplay the scientific importance of robotic Mars sample return, which will in any case be essential before human missions can responsibly be sent to Mars, but merely to point out the step change in sample availability (both in quantity and diversity) that may be expected when and if human missions *are* sent to the planet.

**Bibliometrics**

If, as argued above, human exploration of planetary surfaces is really scientifically so much more efficient than robotic exploration then we would expect to see this reflected in the scientific literature. As the only human exploration missions conducted to-date were the Apollo missions, any such bibliometric comparison must be between publications based on Apollo data and those based on various robotic missions to the Moon and Mars. With this in mind, Figure 2 shows the cumulative number of refereed publications recorded in the SAO/NASA Astrophysics Data System (ADS) resulting from the six successful Apollo landings, the three Luna sample return missions (*Luna*s 16, 20, 24), the two tele-operated Lunokhod rovers (*Luna*s 17, 21), the five successful Surveyor lunar soft landers, and the two MERs (*Spirit* and *Opportunity*) on Mars. Those interested will find more details of the search parameters in the accompanying box.

Several things are immediately apparent from Figure 2. Most obvious is the sheer volume of Apollo's scientific legacy compared to the other missions illustrated. This alone goes a long way to vindicate the points made above about human versus robotic efficiency. The second point to note is that the next most productive set of missions are the lunar sample return missions *Luna*s 16, 20 and 24, which highlights the importance of sample return. Indeed, a large part of the reason why Apollo has resulted in many more publications than the Luna missions is due to the much larger quantity and diversity of the returned samples which, as we have seen, will always be greater in the context of human missions. The third point to note is that, despite being based on data obtained and samples collected over 40 years ago, and unlike the Luna, Lunokhod, or Surveyor publications, which have clearly levelled off, the Apollo publication rate is still rising. Indeed, it is actually rising as fast as, or faster than, the publications rate derived from the Mars Exploration Rovers, despite the fact that data derived from the latter are much more recent. No matter how far one extrapolates into the future, it is clear that the volume of scientific activity generated by the MERs, or other robotic exploration missions, will never approach that due to Apollo.

However, to my mind, the most staggering thing about Figure 2 is that this enormous scientific legacy is based on a total of only 12.5 days total contact time with the lunar surface. Note that this is the total cumulative time the Apollo astronauts were on the Moon, including down time in the Lunar Module, not the cumulative EVA time which was only 3.4 days (Orloff and Harland, 2006). This may be compared with a total of 436 active days on the surface for the Lunokhods (Wilson 1987) and 5162 days for the Mars Exploration Rovers (to the end of 2011, allowing for the fact that contact was lost with *Spirit* on 22 March 2010). This comparison is made graphically in Figure 3, which shows the cumulative number of publications divided by days of fieldwork on the surface (adopting 12.5 days for Apollo to allow for a fair comparison with the rovers). This is the same as dividing the cumulative number of publications by the number of sites studied up to a given date and the average days of fieldwork per site. Dips in the cumulative curves occur when a new mission arrives, instantaneously increasing the days of fieldwork before this has fed through to increased publications. Note the logarithmic scale -- by this metric, Apollo was over three orders of magnitude more efficient in producing scientific papers per day of fieldwork than are the MERs. Coincidentally or otherwise, this is essentially the same as Squyres' (2005) intuitive estimate given above, and is fully consistent the more quantitative analogue fieldwork tests reported by Snook et al. (2007).

**Human exploration costs**

Although it is generally taken for granted that human exploration is more expensive than robotic exploration, and this is certainly true if the aggregate costs are the only ones considered, the situation is not as clear cut as it is sometimes made out to be. For one thing, the ratio of costs between human and robotic missions, while large, may nevertheless be smaller than the corresponding ratio in scientific productivity. The Apollo missions are instructive in this respect. Wilhelms (1993) and Beattie (2001) estimated a total cost of Apollo as $25 billion 'in 1960's money'. This is rather more than the actual Congressional appropriations for Apollo ($19.4 billion from 1961 to 1973; tabulated by Orloff and Harland 2006). Taking the higher estimate (to be conservative) and taking '1960s' to be 1966 when Apollo expenditure peaked, this corresponds to about $175 billion today (where I have made use of the US Bureau of Statistics Inflation Calculator; http://www.bls.gov/data/inflation_calculator.htm).

It is interesting to compare this with the cost of a modern state-of-the-art robotic mission, like Mars Science Laboratory. MSL, which at this writing is en route to Mars, has cost an estimated $2.5 billion (Leone 2011). Thus, in real terms, Apollo cost 70 times as much as MSL. However, Apollo visited six sites, whereas MSL will only visit one site, so in terms of cost per site Apollo was only 12 times as expensive as MSL yet each Apollo mission was vastly more capable. It is true that this comparison only strictly holds in the context of lunar exploration, where we can compare Apollo with a hypothetical future MSL-like lunar rover; in the context of Mars exploration, human missions seem likely to be more expensive than Apollo in real terms (although not necessarily by a large factor -- the estimated total costs of some human Mars mission architectures are comparable to that of Apollo, or even lower; e.g. Turner, 2004). The main point is that human missions like Apollo are between two and three orders of magnitude more efficient in performing exploration tasks than robotic missions, while being only one to two orders of magnitude more expensive. In addition, human missions can accomplish scientific objectives which are unlikely to be achieved robotically at all (deep drilling and properly representative sample collection and return are obvious examples, as well as the increased opportunities for serendipitous discoveries). Looked at this way, human space exploration doesn't look so expensive after all!

That said, there is a more sophisticated and productive way to view the relative costs of human and robotic spaceflight. The fact is that while robotic planetary missions are science-focussed, and essentially their whole costs are therefore borne by scientific budgets, human spaceflight is not wholly, or even mainly, science-driven. Rather, the ultimate drivers of human spaceflight tend to be geopolitical concerns, industrial development and innovation, and employment in key industries. Thus, science can be a beneficiary of human missions instituted and (largely) paid for by other constituencies. Apollo again provides an excellent example. As is well known, Apollo was instituted for geopolitical rather than scientific reasons, and to first order the US Government's expenditure of $25 billion ($175 billion today) would have occurred anyway, whether any science was performed or not. Fortunately, owing largely to the efforts of a relatively small number of senior scientists (documented by Beattie 2001) scientific objectives and capabilities were incorporated into Apollo and, as outlined above, this has resulted in a rich scientific legacy that is still being exploited today.

Of course, including science in Apollo did not come entirely free of cost, and it is interesting to compare this additional, strictly scientific, investment with the cost of robotic missions that were, and are, demonstrably less capable than a human programme like Apollo. Beattie (2001) has conducted a careful study of the cost of including science in Apollo (including the Apollo Lunar Surface Experiment Packages (ALSEPs), grants to the ALSEP investigators, construction of the Lunar Receiving Laboratory in Houston, grants to the initial tranch of lunar sample investigators, astronaut ALSEP and geological training, etc), and arrived at a figure of $350 million in 1972 dollars (somewhat higher than other published estimates). Beattie does not explicitly include the additional cost of developing the Lunar Roving Vehicle ($37 million), but this should probably be added as the LRV was included in the last three Apollo missions mainly to enhance geological exploration. Thus we arrive at a total *scientific* cost in Apollo of $387 million in 1972 dollars. This corresponds to about $2.09 billion today, or 1.2 percent of the total Apollo budget.

**Comparison with robotic exploration costs**

It is instructive to compare this with the costs of some past and planned robotic missions. It is sometimes difficult to get reliable estimates for these, but a search of various internet sources gives $265 million for Mars Pathfinder (which landed on Mars in 1997), $820 million for the two Mars Exploration Rovers (landed in 2004 and considering only the first 90 days of operations), and, as noted above, $2.5 billion for Mars Science Laboratory. Again employing the US Bureau of Statistics Inflation Calculator, these correspond to $374 million, $982 million, and $2.5 billion in 2011 dollars. For comparison, according to a NASA Planetary Sciences Decadal Survey Steering Committee report available on the Web (Li and Hayati 2010), the estimated cost of the proposed Mars Sample Return (MSR) mission, which hopefully will occur around 2025, is about $6.5 billion These mission costs are compared in Figure 4, from which two points are immediately obvious:

(1) The cost of the robotic exploration of Mars has *not* been decreasing as technology advances, but rather has been increasing steadily (Fig. 4). There is a good reason for this -- planetary surfaces are large, rough, rugged places, not at all amenable to exploration with small, cheap rovers no matter how much 'intelligence' can be built into them. There isn't much point in having a hyper-intelligent rover the size of a matchbox if it can only travel 5 meters a day and gets stuck every time it encounters a rock the size of an orange! Nor is such a vehicle likely to carry much in the way of instrumentation. As a consequence, rovers have got bigger and more expensive (and of course more capable) with time (Fig. 5). This is exactly opposite to the trend predicted by some (e.g. Rees 2011), but it is one which will have to continue if we persist in trying to explore planetary surfaces with robots. The problem is, the rising costs may soon be unsustainable within purely scientific budgets

(2) In real terms, the cost of Mars Science Laboratory ($2.5 billion) already exceeds the additional cost of flying science on Apollo ($2.09 billion in today's money). In fact, as Apollo visited six different sites on the Moon, the cost of science per site ($348 million in 2011 dollars) is actually less than the cost of *any* of these robotic Mars missions and, as described above, the scientific efficiency was incomparably higher.

**Conclusions**

The lesson seems clear: if at some future date a series of Apollo-like human missions return to the Moon and/or are sent on to Mars, and if these are funded (as they will be) for a complex range of socio-political reasons, scientists will get more for our money piggy-backing science on them than we will get by relying on dedicated autonomous robotic vehicles which will, in any case, become increasingly unaffordable.

Fortunately, there is a way forward. In 2007 the World's space agencies came together to develop the Global Exploration Strategy (GES), which lays the foundations for a global human exploration programme which could provide us with just such an opportunity (GES 2007). One of the first fruits of the GES has been the development of a Global Exploration Roadmap (GER 2011), which outlines possible international contributions to human missions to the Moon, near-Earth asteroids and, eventually, Mars. The motivations for the GES are, needless-to-say, multifaceted, and include a range of geopolitical and societal motivations (many of them highly desirable in themselves) in addition to science.

From the above discussion, it ought to be clear that science would be a major beneficiary of participating in a human exploration programme such as envisaged by the Global Exploration Strategy. Quite simply, this will result in new knowledge, including answers to fundamental questions regarding the origin and evolution of planets, and the distribution and history of life in the Solar System, that will not be obtained as efficiently, and in many cases probably not obtained at all, by reliance on robotic exploration alone.

**Acknowledgements**
I thank my colleagues Drs Peter Grindrod, Katherine Joy and Heino Falcke, who acted as sounding boards for some of these ideas, and the LPI Librarian, Mary Ann Hager, for advice on collating the Apollo biblometric data. This research has made use of NASA's Astrophysics Data System.

**Wilhelms D E** 1993 *To a Rocky Moon: A Geologist's History of Lunar Exploration* (University of Arizona Press, Tucson).

**Wilson A** 1987 *Solar System Log* (Jane's, London).

-------------------------------

## BOX: Bibliometric search details

The bibliometric statistics shown in Figs 2 and 3 were obtained from the SAO/NASA ADS database. The searches were restricted to refereed publications only. It is likely that these statistics underestimate the total number of publications resulting from these missions, especially for Apollo, as some sample analysis work has been published in geological and petrological journals not normally included in the ADS (although, through the efforts of the library of the Lunar and Planetary Institute (LPI), the ADS should be complete up to about 1995 when the LPI Lunar Bibliography was turned over to the ADS; M.A. Hager, personal communication, 2011). I have made use of the ADS here principally because of its ease of use, and because it is a well-respected bibliometric database in the astronomical community.

The Apollo publications were retrieved by searching for papers containing the words "Apollo" AND "Moon" in the title or abstract (including "Moon" was necessary to exclude papers referring to the Apollo asteroids). These were then examined to check that they do indeed relate to the Apollo missions. Similarly, the Surveyor papers were retrieved by searching for "Surveyor" AND "Moon" (where this time "Moon" was included to exclude references to surveys unrelated to the Surveyor missions). The Luna 16/20/24 and Lunokhod publications were retrieved by searching for these missions by name. Finally, the MER publications were retrieved by searching for [(("Spirit" OR "Opportunity") AND ("rover" OR "Gusev" OR "Meridiani")) OR "Mars Exploration Rover"]; this slightly complicated set of search criteria was rendered necessary to avoid the surprisingly large number of papers reporting work conducted in such and such a "spirit" or which provide great "opportunities" for various things unrelated to Mars exploration! A check was made against the MER Science Team's own publication list (http://marsrover.nasa.gov/science/pdf/web_publist.pdf) to ensure that these papers were correctly recovered by the ADS search criteria (although the Team papers are only a subset of the total number of papers making reference to the MERs included in Fig. 2, this agreement gives confidence that a large number of MER papers have not been missed in the ADS search).

Doubtless more sophisticated bibliometric analyses could be performed, and this might be an interesting exercise for someone, but the results shown in Figures 2 and 3 appear sufficient for present purposes.

-------------------------

## FIGURES

| Skill | Objective Measurement | Humans | Advantage | Robots |
|---|---|---|---|---|
| • Strength | Y | High strength/high torque; sometimes too strong | 🟢 | Load/torque can be varied over very wide range with precise control |
| • Endurance | Y | Limited by available consumable and physical tolerances | 🟢 | Limited by design, environmental decay |
| • Precision | Y | High degree of training is required to ensure repeated performance in humans | 🔴 | Once programmed, robot precision is limited only by electromechanical design |
| • Cognition | N | Creative and limited only by prior training | 🟢 | Execution of pre-programmed routines |
| • Perception | N | Highly integrated sensory suite of limited use in space environment, visual acuity is very high | 🟢 | Can detect minute environmental changes; can sense trace elements in low concentration |
| • Detection | Y | High detection sensitivity, though sensory paths limited during exploration | 🟢 | Extremely high detection ability if preprogrammed and equipped with proper sensors |
| • Sensory Accuity | Y | Highly integrated sensory suite of limited use in space environment, visual acuity is very high | 🔴 | Capable of detecting minute environmental changes if so equipped |
| • Speed | Y | Able to cover great distances in short amount of time | 🟢 | Able to work very slowly and steadily |
| • Response time | Y | Spot decisions and rapid response is customary, sometimes a disadvantage | 🟢 | Rapid to programmed events, latency delay for "hold" events |
| • Decision making | N | Flexible, unlimited in either speed or capacity | 🟢 | Primitive learning capacity to scripted events |
| • Reliability | Y | High in terms of meeting mission objectives but require support systems of high complexity | 🔴 | High reliability, but relatively short lives in space exploration environments |
| • Adaptability | N | Highly adaptable to new and changing situations | 🟢 | Reprogrammable to a limited extent, otherwise limited by design and system redundancy |
| • Agility | N | Agility limited only by design of exoshell | 🟢 | Computation requirements dictate slow movements with limited agility |
| • Versatility | N | Readily self-reprogrammable to provide multi-purpose services and functions | 🟢 | Generally designed to perform specific functions and poorly equipped to new applications |
| • Dexterity | Y | Ability to manipulate large and very small objects with high flexibility | 🟢 | Can exhibit very high DoF and fast reaction times |
| • Fragility | N | Generally robust but total system failure can be caused by small affects | 🟢 | Exploration robots generally very fragile, especially attendant instrument suites |
| • Expendibility | N | Human life is precious and we place ever higher value on it | 🔴 | Robot also high value - today we send robots to less interesting scientific sights on Mars because value is high - return unnecessary |
| • Maintainability | N | Low cycle time between periodic consumable replenishments, requires expert skills to maintain | 🟡 | Limited only by design - can be maintained by low skill personnel |

**Figure 1.** Tabular summary of exploration skills reported by Garvin (2004). The fourth column indicates relative advantage of humans or robots on a sliding scale: green symbols indicate that the balance of advantage lies with humans, red with robots, and yellow that both are approximately equal. In most cases humans have a clear advantage. Since 2004 the extreme endurance of the MERs has moved this entry more in favour of robots, but they remain slow and inflexible. (Figure courtesy of Jim Garvin/NASA/Springer, and reproduced with permission.)

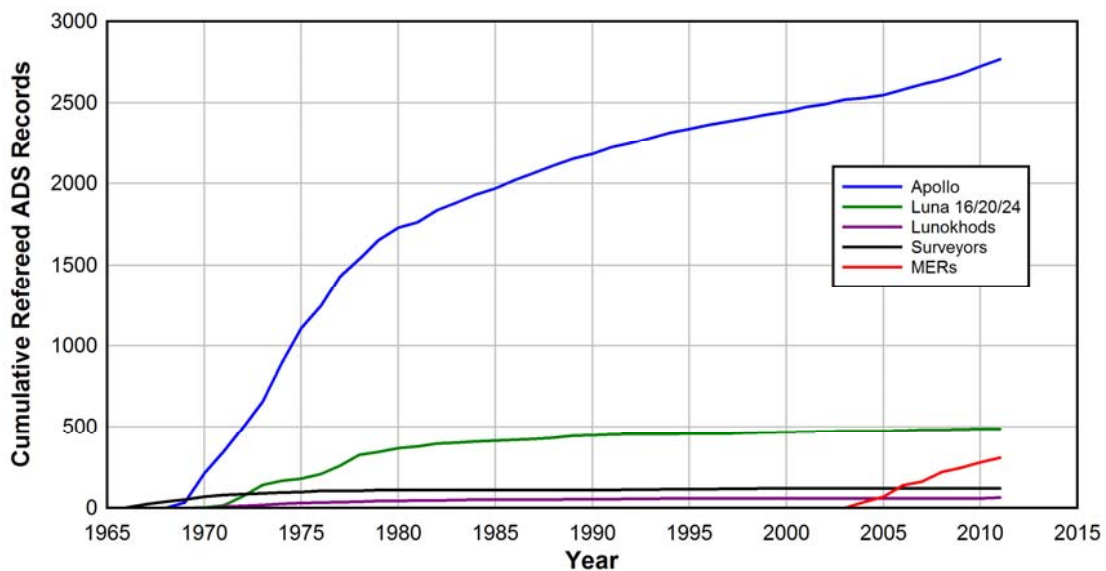

**Figure 2.** Cumulative number of refereed publications in the ADS database for the Apollo, Luna 16/20/24, Lunokhod, Surveyor, and Mars Exploration Rover missions (see Box for details).

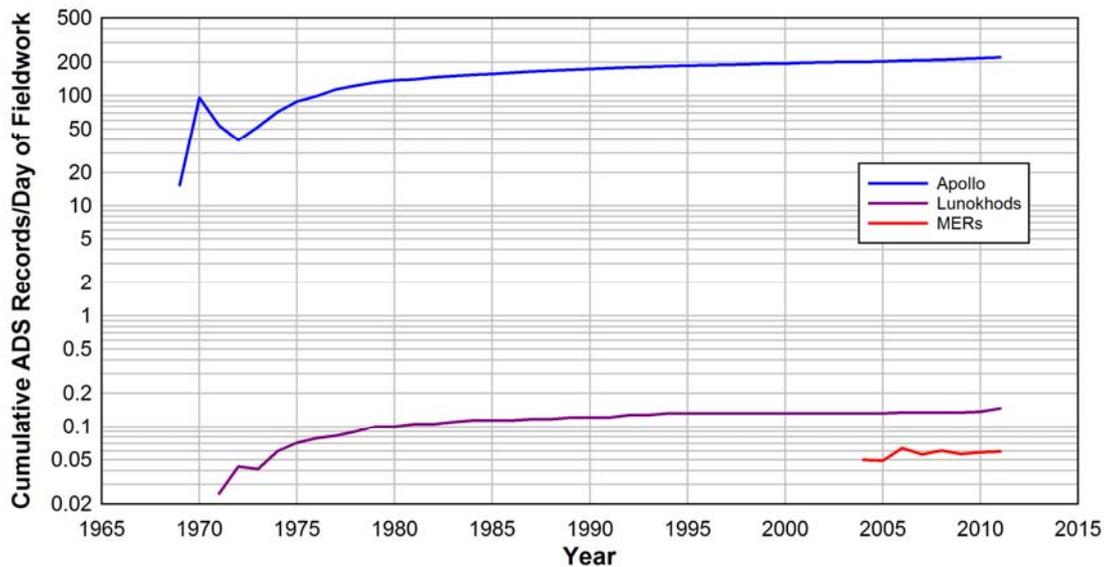

**Figure 3.** Cumulative number of refereed publications per day exploring the surface of the Moon or Mars. This is the same as dividing the cumulative number of publications by the number of sites visited by a given date and the average time spent at those sites. The plot is restricted to those missions plotted in Fig. 2 which had, or have, surface mobility, and which can therefore be considered as having performed 'fieldwork'.

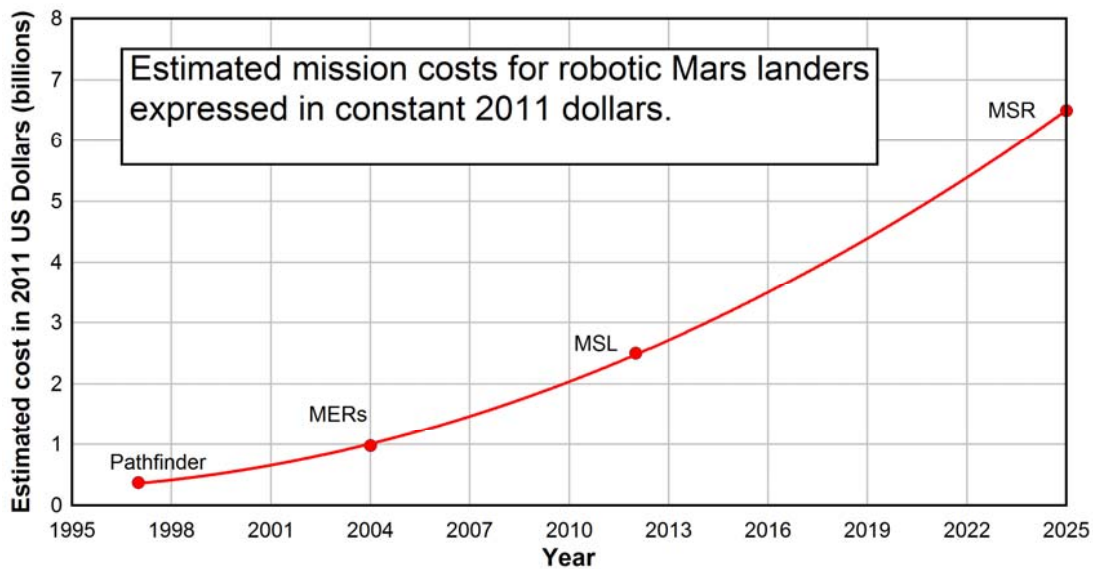

**Figure 4.** Estimated costs of robotic Mars rover missions expressed in constant 2011 dollars. Note that there were two MERs so the individual cost of each would be less than the plotted value (but perhaps not by much given the economies of scale inherent in producing two rather than one); as currently conceived, MSR requires two rovers (possibly including ESA's ExoMars) in addition to other expensive elements). The curve is a 2nd order polynomial fit to the trend – for how much longer will this be sustainable within purely scientific budgets? Note that this is not an inflationary increase (real-terms costs are plotted): we could continue to build rovers as small and (relatively) cheap as Pathfinder, but choose not to owing to their inherent scientific limitations.

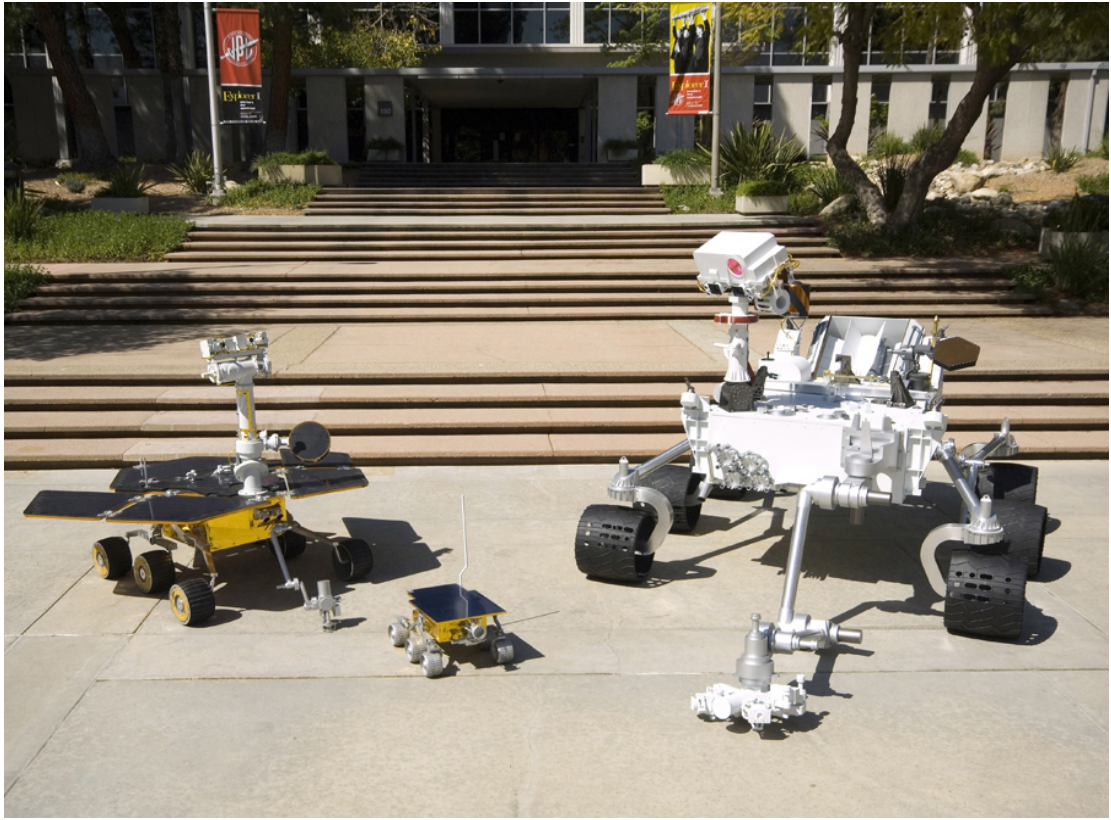

**Figure 5.** The increasing size of Mars rovers, from Pathfinder (centre), the MER's (left), to MSL (right). This increase in size (and cost), which is exactly opposite to predictions that improved technology will result in smaller and cheaper robots, is mandated by the nature of the Martian surface and complexity of exploration objectives. Human missions would be even larger and more expensive, but, crucially, much more capable (NASA/JPL).